\PassOptionsToPackage{dvipsnames}{xcolor}
\documentclass[twocolumn,trackchanges]{aastex631}
\usepackage{newtxtext,newtxmath}
\usepackage[T1]{fontenc}
\usepackage{anyfontsize}
\usepackage{threeparttable}
\usepackage{booktabs}
\usepackage{multirow}
\usepackage{graphicx}
\usepackage{amsmath}
\usepackage{float}
\newcommand{\RomanNumeralCaps}[1]{\MakeUppercase{\romannumeral #1}}
\received{August 14, 2024}
\revised{September 20, 2024}
\accepted{October 8, 2024}
\shorttitle{{\sc 21cmLSTM}: memory-based 21 cm emulation} \shortauthors{Dorigo Jones et al.} \graphicspath{{./}{}}
\makeatletter
\newcommand\footnoteref[1]{\protected@xdef\@thefnmark{\ref{#1}}\@footnotemark}
\makeatother

\begin{document}
\title{{\sc 21cmLSTM}: A Fast Memory-based Emulator of the Global 21 cm Signal with Unprecedented Accuracy}
\author[0000-0002-3292-9784]{J. Dorigo Jones} \affiliation{Center for Astrophysics and Space Astronomy, Department of Astrophysical and Planetary Sciences, University of Colorado Boulder, CO 80309, USA}
\author[0000-0003-0016-5377]{S. M. Bahauddin} \affiliation{Laboratory for Atmospheric and Space Physics, University of Colorado, Boulder, CO 80303, USA}
\author[0000-0003-2196-6675]{D. Rapetti} \affiliation{NASA Ames Research Center, Moffett Field, CA 94035, USA} \affiliation{Research Institute for Advanced Computer Science, Universities Space Research Association, Washington, DC 20024, USA} \affiliation{Center for Astrophysics and Space Astronomy, Department of Astrophysical and Planetary Sciences, University of Colorado Boulder, CO 80309, USA}
\author[0000-0002-8802-5581]{J. Mirocha} \affiliation{Jet Propulsion Laboratory, California Institute of Technology, 4800 Oak Grove Drive, Pasadena, CA 91109, USA} \affiliation{California Institute of Technology, 1200 E. California Boulevard, Pasadena, CA 91125, USA}
\author[0000-0002-4468-2117]{J. O. Burns} \affiliation{Center for Astrophysics and Space Astronomy, Department of Astrophysical and Planetary Sciences, University of Colorado Boulder, CO 80309, USA}
\correspondingauthor{J. Dorigo Jones} \email{johnny.dorigojones@colorado.edu}
\begin{abstract}
Neural network (NN) emulators of the global 21 cm signal need emulation error much less than the observational noise in order to be used to perform unbiased Bayesian parameter inference. To this end, we introduce {\tt 21cmLSTM} -- a long short-term memory (LSTM) NN emulator of the global 21 cm signal that leverages the intrinsic correlation between frequency channels to achieve exceptional accuracy compared to previous emulators, which are all feedforward, fully connected NNs. LSTM NNs are a type of recurrent NN designed to capture long-term dependencies in sequential data. When trained and tested on the same simulated set of global 21 cm signals as the best previous emulators, {\tt 21cmLSTM} has average relative rms error of 0.22\% -- equivalently 0.39 mK -- and comparably fast evaluation time. We perform seven-dimensional Bayesian parameter estimation analyses using {\tt 21cmLSTM} to fit global 21 cm signal mock data with different adopted observational noise levels, $\sigma_{21}$. The posterior $1\sigma$ rms error is $\approx3\times$ less than $\sigma_{21}$ for each fit and consistently decreases for tighter noise levels, showing that {\tt 21cmLSTM} can sufficiently exploit even very optimistic measurements of the global 21 cm signal. We made the emulator, code, and data sets publicly available so that {\tt 21cmLSTM} can be independently tested and used to retrain and constrain other 21 cm models.
\end{abstract}
\section{Introduction} \label{sec:intro}
Neutral hydrogen (HI) emits radiation at 1420.4 MHz ($\lambda\approx$21 cm) via the spin-flip transition that coupled with the gas kinetic temperature during the Epoch of Reionization (EoR; ending by $z\approx6$), Cosmic Dawn (CD; $10\lesssim z \lesssim30$), and Dark Ages ($z>30-40$; for reviews see \citealt{Furlanetto06, Bera23}). As a result, the differential brightness temperature of the 21 cm line with respect to the cosmic microwave background (CMB), $\delta T_b$, is expected to be a powerful probe of the astrophysics and cosmology of each of these cosmic epochs. Numerous low-frequency radio experiments ($\nu\lesssim225$ MHz, corresponding to $z\gtrsim5.3$) have pursued measurements of the sky-averaged (i.e., global; \citealt{Shaver99}) 21 cm signal \citep{EDGES, SARAS2, SARAS3, REACH}, as well as its power spectrum \citep{GMRT, LOFAR, MWA, LEDA, HERA}, although systematic effects, mainly from the galactic foreground in combination with beam chromaticity and radio frequency interference (RFI), have so far prevented a clear detection of the global signal (e.g., \citealt{Hills18, Bradley19, SimsPober20, Tauscher20}). Efforts are also underway to measure the Dark Ages 21 cm signal from the Moon (e.g., LuSEE-Night, \citealt{Bale23}).

To constrain physical parameters able to describe the global 21 cm signal, Bayesian inference is a powerful tool (e.g., \citealt{Bernardi16, Liu20, PaperII, Shen22}). Likelihood-based inference techniques such as Markov Chain Monte Carlo (MCMC) and nested sampling (\citealt{Skilling04}) are used to numerically estimate the parameters of models from data and constrain the full joint posterior distribution (e.g., \citealt{Schmit18, Mirocha19, Monsalve19, Bevins22a, Bevins24}). Bayesian inference can require $10^6$ or more model evaluations to fully search the prior volume and calculate the posterior, which can become exceedingly computationally expensive, especially when constraining many parameters and jointly fitting for different systematics.

Machine learning in the form of artificial neural networks can be employed to mimic the physical models of interest being sampled in a Bayesian fitting analysis and efficiently obtain converged posteriors.~Through supervised learning of labeled data generated by the physical model, networks can be taught the relationship between the input parameters and the output (in this case, $\delta T_b$) to quickly and accurately emulate the model. Emulation error on the order of 1 mK can result in significantly biased posteriors even when fitting global 21 cm signal mock data with statistical noise of 25 mK \citep{DorigoJones23}, and so emulation error $<1$ mK is needed to sufficiently exploit optimistic or standard measurements of the 21 cm signal and obtain unbiased posteriors.

Long short-term memory (LSTM; \citealt{HochreiterSchmidhuber97, Gers00})  networks are a type of recurrent neural network (RNN), which differ from feedforward networks such as fully connected neural networks (FCNNs; e.g., \citealt{Rumelhart86}), also called multilayer perceptrons, and convolutional neural networks (CNNs). In short, FCNNs have a one-directional flow of information that is not specifically designed for temporal awareness, while RNNs have feedback connections that allow them to learn trends or sequences of features in data (e.g., \citealt{LeCun15}). LSTM networks have been successful in numerous temporal prediction and classification problems in astrophysics (\citealt{Liu19, Hu22, Sun22, Iess23, Zheng23, Tabasi23, Huber24, Li24}), although their ability to emulate a physical, numerical model is relatively unexplored \citep{Zhang20}.

So far, LSTM networks or RNNs have not been utilized to emulate the global 21 cm signal or any summary statistic in 21 cm cosmology. At the time of writing this paper, there exist four publicly available, neural-network-based emulators of the global 21 cm signal -- {\sc 21cmGEM} \citep{Cohen20}, {\tt globalemu} \citep{globalemu}, {\tt 21cmVAE} \citep{21cmVAE}, and {\tt 21cmEMU} \citep{Breitman24} -- all of which use FCNNs to predict $\delta T_b$ as a function of the independent variable, being redshift or frequency, given (seven to nine) input astrophysical parameters. In this paper, we present a novel LSTM-based emulator of the global 21 cm signal, called {\tt 21cmLSTM}, which exploits the intrinsic correlation of information between adjacent frequency channels (i.e., autocorrelation) in 21 cm data to achieve unprecedented emulation accuracy. \citet{Prelogovic22} found similar benefit of LSTM RNNs but as a regressor for 21 cm 3D lightcones, when used with convolutional layers (see \citealt{Shi15, KodiRamanah22}).

For detailed descriptions of RNNs and LSTM cells, see, e.g., \citet{StaudemeyerRothsteinMorris19} and \citet{Sherstinsky20}; here we provide a conceptual overview. The ``hidden state'' is the key element of RNNs, which reuses the same weights and biases on each step and updates them via back propagation through time (BPTT; \citealt{Williams90}). Information is fed through the RNN sequentially, and the hidden state output from each step is used to inform the output of all future steps. Basic RNNs are limited to predicting $\sim$10 time steps, though, because of the ``vanishing gradient'' problem, whereby the backpropagated error either vanishes or explodes as more weights are multiplied together (e.g., \citealt{Pascanu13}). LSTM cells were invented to avoid this problem by incorporating a ``memory cell internal state,'' or ``information highway,'' that enforces constant error flow. LSTM cells contain forget, input, and output gates that determine the relative importance of each time step and ensure the gradient can bridge 1000 or more steps without vanishing, thereby helping to identify both short-term and long-term correlations in data. For a single-layer (i.e., non-stacked) LSTM network, the number of activation operations is the data resolution (i.e., number of channels or bins), and so hyperparameter optimization relies purely on determining the best number of layers of nonlinear activation and the number of training epochs, whereas FCNNs contain an additional dimension to optimize, being the number of nodes per hidden layer.

The paper is organized as follows: In Section~\ref{sec:methods}, we describe the architecture and training of {\sc 21cmLSTM}; in Section~\ref{sec:results}, we present the emulation accuracy and speed of {\sc 21cmLSTM}; in Section~\ref{sec:posterioremulation}, we present the posterior constraints when using {\sc 21cmLSTM} in a Bayesian nested sampling analysis fitting mock data; and in Section~\ref{sec:conclusions}, we summarize the conclusions.

\section{Methods} \label{sec:methods}
In this section, we describe the components of {\sc 21cmLSTM}, including the network architecture, the data sets used for training, validation, and testing, the data preprocessing steps, the training settings, and the optimization performed to ensure robust and accurate emulation results. The emulator is written in {\sc Python} using the {\sc Keras} \citep{keras} machine learning libraries with {\sc TensorFlow} \citep{tensorflow} backend. The code\footnote[7]{\label{code}\url{https://github.com/jdorigojones/21cmLSTM}} and data\footnote[8]{\label{data}\href{https://doi.org/10.5281/zenodo.5084113}{DOI: 10.5281/zenodo.5084113}; \href{https://doi.org/10.5281/zenodo.13840725}{DOI: 10.5281/zenodo.13840725}} are both publicly available, making {\sc 21cmLSTM} simple to use and retrain.

\subsection{Architecture} \label{subsec:architecture}
\begin{figure}
    \includegraphics[scale=0.47]{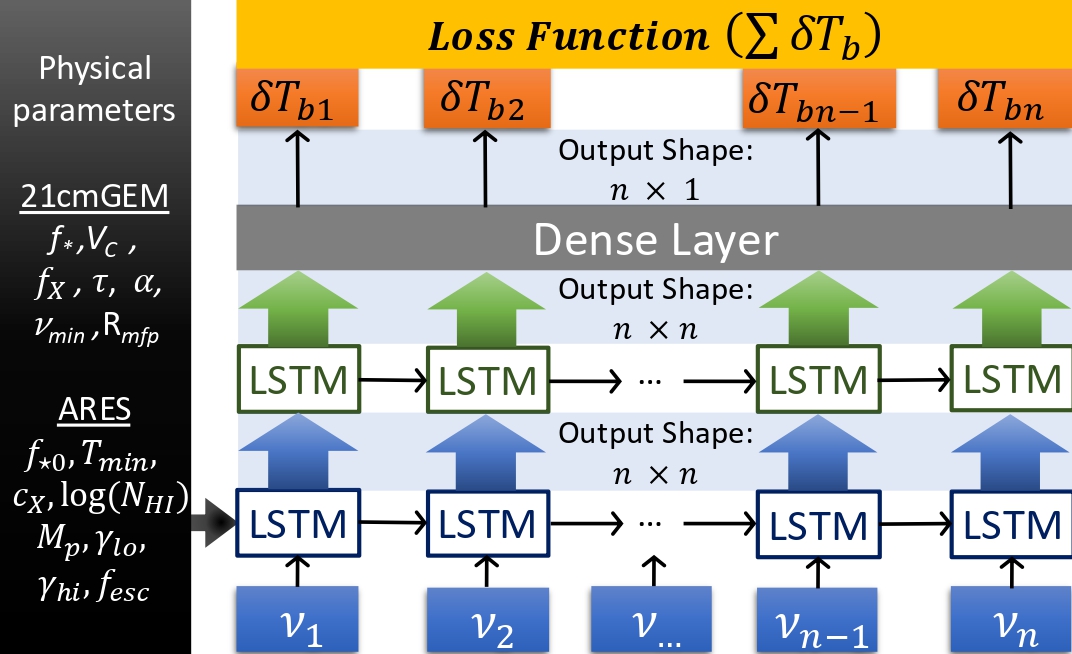}
    \caption{Schematic diagram of {\sc 21cmLSTM} network architecture. The user inputs the physical parameter values for the desired model, and the emulator predicts $\delta T_b$ for all frequencies. Arrows indicate inputs to or outputs of layers and LSTM cells. The input array is $(N,n,p)$, where $N$ is the number of signals, $n$ is the number of frequency channels per signal, and $p$ is the number of physical parameters plus one for the frequency channel. For emulating the {\sc 21cmGEM} and {\tt ARES} training sets, $(N,n,p)$ is (24,562, 451, 8) and (23,896, 449, 9), respectively. See Section~\ref{subsec:architecture} for further details.}
    \label{fig:emulator}
\end{figure}

The emulator model is composed of two LSTM layers (i.e., two layers of nonlinear activation, equivalent to two hidden layers in a FCNN), followed by a dense layer with output dimensionality of one. Figure~\ref{fig:emulator} shows a schematic diagram of the {\sc 21cmLSTM} network architecture with arrows indicating connections between layers or between LSTM cells. The emulator takes as input the physical parameters, which are user-defined, along with the list of frequencies, which is initialized within the emulator, and outputs the brightness temperature, $\delta T_b$, for all frequencies. The emulator creates a 3D input array, $(N,n,p)$, for the first LSTM layer, where $N$ is the number of signals, $n$ is the number of frequency channels in each signal, and $p$ is the number of physical parameters plus one for the frequency channel. The LSTM cells are ``many-to-many,'' meaning each cell predicts the entire signal sequence, and so each LSTM layer has output dimensionality equal to the number of frequency channels. As mentioned, the second LSTM layer is connected to a fully-connected output layer that predicts $\delta T_b$ for each frequency channel, which is used to calculate the loss during BPTT.

The LSTM layers use hyperbolic tangent ($\tanh$) activation function, and the output layer uses linear activation. The model uses the Adam stochastic gradient descent optimization method \citep{Adam} and mean squared error (MSE) loss function:
\begin{equation} \label{eqn:mse}
{\rm MSE} = \langle(\delta T_b(\nu)-\hat{\delta T_b}(\nu))^2\rangle,
\end{equation}

\noindent where $\hat{\delta T_b}(\nu)$ is the emulated signal produced by {\sc 21cmLSTM}, and $\delta T_b(\nu)$ is the simulated, ``true'' signal produced by the model on which the emulator is trained. We performed ``hyper-parameter tuning'' to minimize prediction error, by testing one LSTM layer, three LSTM layers, different activation functions, and mean absolute error loss function, and found the choices stated above result in the most accurate network on average. We also trained a one-layer Bidirectional LSTM (Bi-LSTM) model, which is an LSTM network trained in both directions, and found that it performs only slightly worse than the two-layer LSTM.

\subsection{Data Sets} \label{subsec:data}
We train and test {\sc 21cmLSTM} on the exact same publicly available set\footnoteref{data} of global 21 cm signals used to originally train and test the previous emulators {\sc 21cmGEM}, {\tt globalemu}, and {\tt 21cmVAE}. The data set was created (see \citealt{Cohen20} for description) by a seminumerical model \citep{Visbal12, Fialkov13, Fialkov14} that is similar to {\sc 21cmFAST} (\citealt{Mesinger11}). We refer to this data set as the {\sc 21cmGEM} set, in which seven astrophysical parameters are varied (Table~\ref{tab:params}; see \citealt{Cohen20}), and each signal spans the redshift range $z=5-50$ with resolution $\delta z=0.1$. Parameter range restrictions and observational constraints are applied, as stated in \citet{Cohen20}, \citet{globalemu}, and \citet{21cmVAE}, to obtain 24,562 training signals, 2,730 validation signals, and 1704 test signals. A representative subset of the {\sc 21cmGEM} combined training+validation set is shown in the top left panel of Figure~\ref{fig:big}.

We also train and test {\sc 21cmLSTM} on a different data set generated by another popular model for the global 21 cm signal, Accelerated Reionization Era Simulations ({\tt ARES}\footnote[9]{\url{https://github.com/mirochaj/ares}}; \citealt{Mirocha14, Mirocha17}), which is a physically-motivated, semianalytical code that is the union of a 1D radiative transfer code \citep{Mirocha12} and a uniform radiation background code \citep{Mirocha14}. We created the {\tt ARES} set to be nearly equivalent to the {\sc 21cmGEM} set, in order to directly compare the accuracy of {\sc 21cmLSTM} between the two models, with: (i) the same size of the test set (1704), and similar (to within 3\%) size of the combined training+validation set (26,552, also split 90\% for training and 10\% for validation); (ii) eight (instead of seven) astrophysical parameters varied over wide ranges (see Table~\ref{tab:params}), which also control the star formation efficiency (SFE) and ionizing photon production in galaxies, although via a different parameterization (see Section~\ref{subsec:model_comparison} for a comparison between the two sets); (iii) the same redshift resolution and nearly identical range ($z=5.1-49.9$); and (iv) a similar physical EoR constraint on the neutral hydrogen fraction ($x_{\rm HI}$) at $z<6$; we require $x_{\rm HI}<$5\% at $z=5.3$, while the {\sc 21cmGEM} set requires $x_{\rm HI}<$16\% at $z=5.9$ based on less recent constraints (see, e.g., \citealt{Fan06, McGreer15, Mason19, Zhu22, Bosman22, Jin23}). A representative subset of the {\tt ARES} combined training+validation set is shown in the bottom left panel of Figure~\ref{fig:big}.

Before the emulator is trained, the training and validation data are preprocessed to be normalized between zero and one, which is usual to facilitate network performance. Some of the physical parameters are uniform only in $\log10$-space, and so the $\log10$ is taken of these parameters: $f_X$, $V_c$, and $f_{\star}$ for {\sc 21cmGEM}, and $c_X$, $T_{\rm min}$, $f_{\rm \star,0}$, and $M_{\rm p}$ for {\tt ARES}. We note that $f_X = c_X/2.6\times10^{39} {\rm erg\,s}^{-1}(M_{\odot} {\rm yr}^{-1})^{-1}$. The signals (i.e., $\delta T_b$ labels and frequency list) are flipped so that the network is trained from high-$z$ to low-$z$. Finally, we performed a Min-Max normalization (Equation~\ref{eqn:minmax}) on each feature, $x$, in the data (i.e., physical parameter values and the list of frequencies) and labels (i.e., $\delta T_b$):
\begin{equation} \label{eqn:minmax}
\tilde{x} = \frac{x-x_{\rm min}}{x_{\rm max}-x_{\rm min}}.
\end{equation}

\noindent We found that normalizing the labels bin-by-bin per signal caused the preprocessed signals to blow up at frequencies with little variation (i.e., small denominator in Equation~\ref{eqn:minmax}). Therefore, the Min-Max normalization is performed globally for the signal labels in order to preserve their original, smooth shape, and for consistency the same is done when normalizing the data. 

The emulator is trained on the preprocessed training set signals and saves at each epoch the MSE loss of the network evaluated on the training and validation sets. Only the training set errors are used during BPTT to update the network weights, while the validation set is used to gauge the emulator's ability to generalize to unseen signals and to check for overfitting. The test set, which is created separately from the training and validation sets, determines the ultimate accuracy of the trained instance of {\sc 21cmLSTM}.

\begin{table*}
    \caption{Astrophysical Parameters Varied in {\sc 21cmGEM} and {\tt ARES} Data Sets and Fit in Nested Sampling Analyses \label{tab:params}}
    \begin{center}
    \begin{tabular}{cccc}
    \toprule
    Model & Parameter & Description & Range (with units)\\
    \hline
    \multirow{7}{*}{{\sc 21cmGEM}} & $f_*$ & star formation efficiency & Log unif. [$10^{-4}$, $5\times10^{-1}$]\\
    & $V_c$ & minimum circular velocity of star-forming halos & Log unif. [$4.2$, $100$] km s$^{-1}$\\
    & $f_X$ & X-ray efficiency of sources & Log unif. [$10^{-6}$, $10^3$]\\
    & $\tau$ & CMB optical depth & Uniform [$0.04$, $0.2$]\\
    & $\alpha$ & slope of X-ray spectral energy distribution (SED) & Uniform [$1$, $1.5$]\\
    & $\nu_{\rm min}$& low energy cut off of X-ray SED & Uniform [$0.1$, $3$] keV\\
    & $R_{\rm mfp}$& mean free path of ionizing radiation &  Uniform [$10$, $50$] Mpc\\    
    \cline{1-4}
    \multirow{8}{*}{{\tt ARES}} & $f_{\rm \star,0}$ & peak star formation efficiency & Log unif. [$10^{-5}$, $10^0$] \\
    & $T_{\rm min}$ & minimum temperature of star-forming halos & Log unif. [$3\times 10^2$, $5\times 10^5$] K \\
    & $c_X$ & normalization of $L_X$--SFR relation & Log unif. [$10^{36}$, $10^{44}$] erg s$^{-1}$($M_{\odot}$ yr$^{-1}$)$^{-1}$\\
    & $\log N_{\rm H \RomanNumeralCaps{1}}$ & neutral hydrogen column density in galaxies & Uniform [18, 23] \\
    & $M_{\rm p}$ & dark matter halo mass at $f_{\rm \star,0}$ & Log unif. [$10^8$, $10^{15}$] $M_{\odot}$ \\
    & $\gamma_{\rm lo}$ & low-mass slope of $f_{\rm \star} (M_{\rm h})$& Uniform [0, 2] \\
    & $\gamma_{\rm hi}$ & high-mass slope of $f_{\rm \star} (M_{\rm h})$ & Uniform [-4, 0] \\
    & $f_{\rm esc}$ & escape fraction of ionizing radiation & Uniform [0, 1] \\
    \bottomrule
    \end{tabular}
    \end{center}
\end{table*}

\subsection{Training} \label{subsec:training}
For results presented in this work, we trained and tested {\sc 21cmLSTM} using a single NVIDIA A100 GPU with 32 CPU cores on the Blanca shared ``condo'' compute cluster operated by University of Colorado Research Computing. The emulator is trained first for 75 epochs with batch size of 10 (i.e., training on batches of 10 signals at a time), then for 25 epochs with batch size of one, then finally for another 75 epochs with batch size of 10. The final saved network loads the model weights and biases from the final epoch of training. Training with a large batch size before and after a smaller batch size (i.e., batch size scheduling; see \citealt{Smith17}) is an increasingly common alternative to decaying the learning rate that facilitates robust gradient descent and speeds up the overall training time, which is an average of 12.4 hr $\pm$ 0.1 hr (utilizing $\approx$6 GB of memory) when training {\sc 21cmLSTM} on the {\sc 21cmGEM} set. We tested different batch sizes between 1 and 32 and found the ones stated above to ensure the model learns efficiently, generalizes well to unseen data, and makes effective use of computational resources.

Instead of incorporating an early stopping condition for the training, we determined the approximate number of training epochs that produces the most accurate and robust resulting network on average. We trained and tested {\sc 21cmLSTM} on the {\sc 21cmGEM} set for 20, 25, 30, and 40 epochs of batch size one, running six trials for each, and find that they have average relative rms errors (see Equation~\ref{eqn:rel_error} below) of 0.34\%, 0.24\%, 0.44\%, and 0.29\%, respectively. We performed this testing with both 75 epochs and 100 epochs of batch size 10 before and after the epochs of batch size one, and we find marginal difference between the two. Therefore, since 25 epochs of training with batch size one, with 75 epochs of batch size 10 before and after, produced the most accurate trained {\sc 21cmLSTM} and with no spurious outlier trials, we employ this training epoch configuration. We find that the validation loss curves reach a stable solution near the end of training (see Figure~\ref{fig:loss}), rather than increasing, which indicates that there is no overfitting.

\section{Emulation Results} \label{sec:results}
\subsection{Accuracy} \label{subsec:accuracy}
\begin{figure*}
    \includegraphics[scale=0.59]{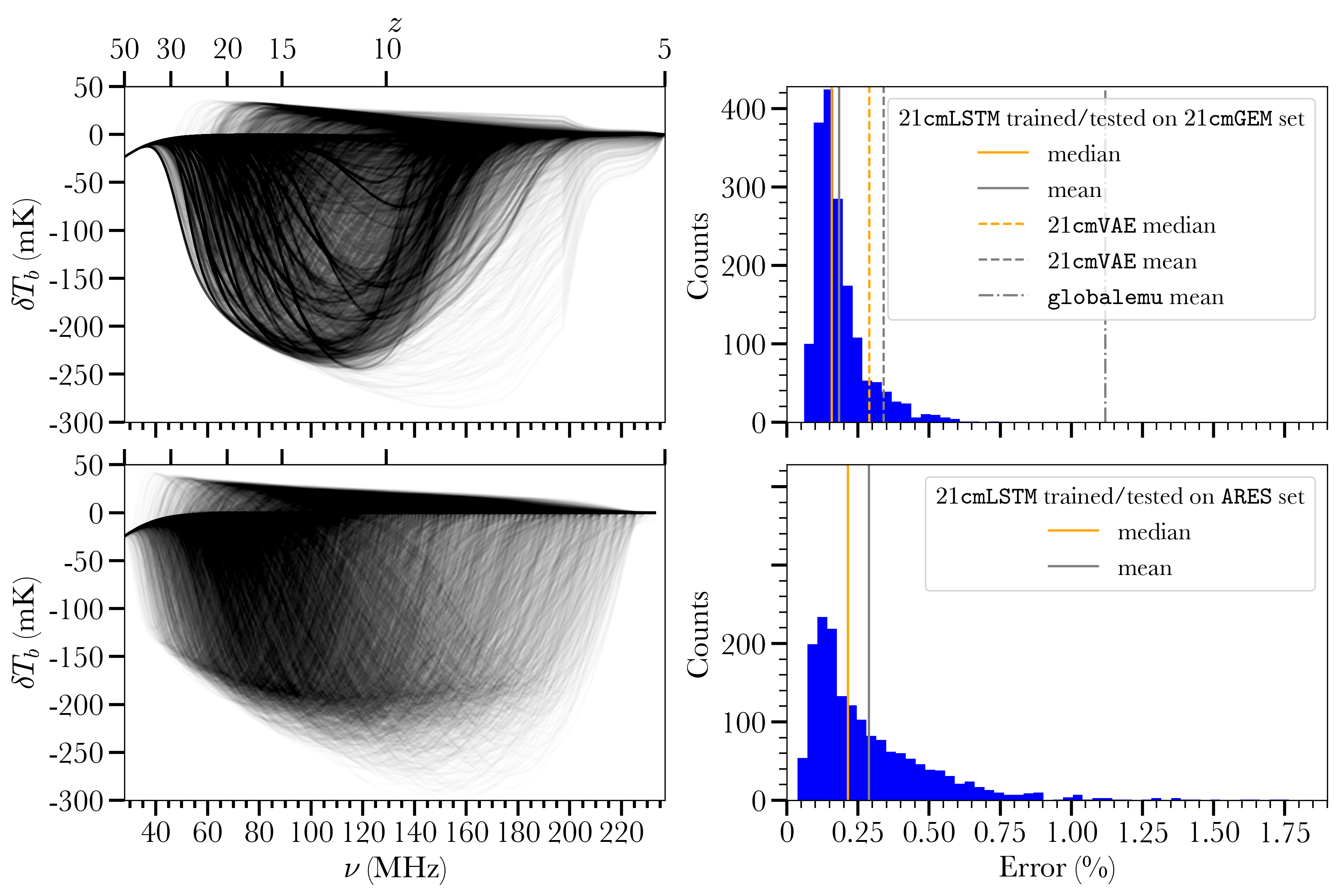} \caption{{\it Left:} Model realizations of 10,000 global 21 cm signals randomly drawn from the {\sc 21cmGEM} (top) and {\tt ARES} (bottom) combined training+validation sets (Section~\ref{subsec:data}). The {\tt ARES} set has $1.3\times$ more statistical outliers and higher PCA error (see Section~\ref{subsec:model_comparison}), which is consistent with more signal variation and consequently larger emulation error (Section~\ref{subsec:accuracy}). {\it Right:} Histograms of the relative rms error (Equation~\ref{eqn:rel_error}) for the best trial of {\sc 21cmLSTM} trained on the {\sc 21cmGEM} (top) and {\tt ARES} (bottom) training sets and evaluated on the 1704 signals in each test set. Vertical gray (orange) lines depict the mean (median) error for {\sc 21cmLSTM} (solid), {\sc 21cmVAE} (dashed), and {\tt globalemu} (dash-dotted).} \label{fig:big}
\end{figure*}

\begin{figure}
    \includegraphics[scale=0.59]{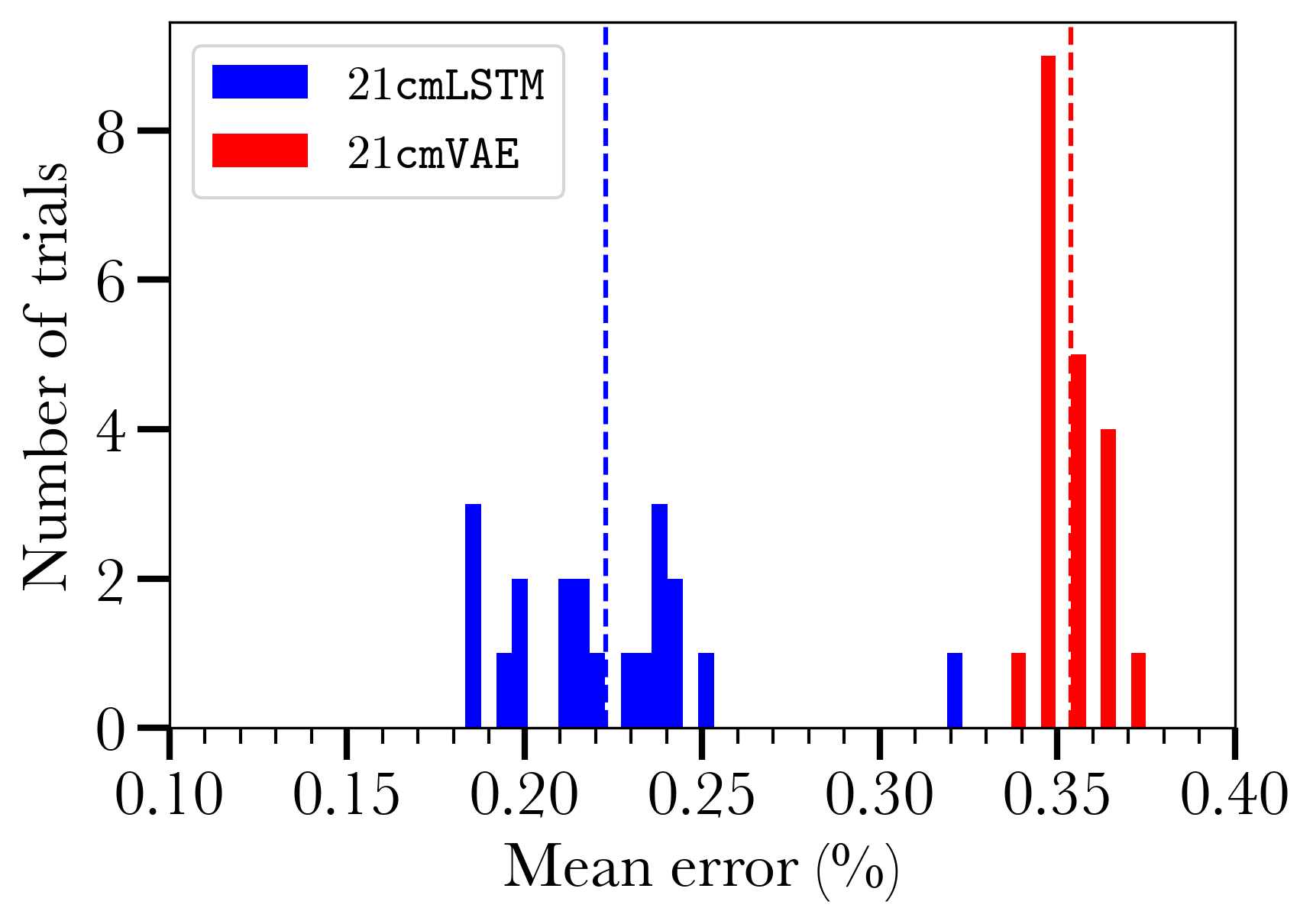} \caption{Histogram of mean relative error for 20 trials of {\sc 21cmLSTM} (in blue) trained and tested on the {\sc 21cmGEM} data set. The red histogram is the approximate error for 20 trials of {\sc 21cmVAE} trained and tested on the same data, adapted from Figure 6 of \citet{21cmVAE}. The dashed blue (red) line depicts the average error for {\sc 21cmLSTM} ({\sc 21cmVAE}).}
    \label{fig:err_hist}
\end{figure}

We report the emulation accuracy of {\sc 21cmLSTM} when trained and tested on the same sets of global 21 cm signals that were used in the original papers for {\sc 21cmVAE} \citep{21cmVAE}, {\tt globalemu} \citep{globalemu}, and {\sc 21cmGEM} \citep{Cohen20}, allowing a direct comparison to these existing emulators (see Table~\ref{tab:results}). We also train and test {\sc 21cmLSTM} on an equivalent data set created by {\tt ARES} (see Section~\ref{subsec:data}).

Using the optimized network architecture and training settings described in Section~\ref{sec:methods}, we trained 20 identical trials of {\sc 21cmLSTM} on the {\sc 21cmGEM} set in order to characterize the stochasticity of the training algorithm. We evaluated each trained network at the parameter values of the 1704 signals in the {\sc 21cmGEM} test set and compared the resulting emulated signals to their corresponding ``true'' signals, computing for each signal the rms error across the full frequency range in both absolute units (mK), and relative units (per cent) as:
\begin{equation} \label{eqn:rel_error}
{\rm Error} = \frac{\sqrt{{\rm MSE}}}{{\rm max}(|\delta T_b(\nu)|)},
\end{equation}
\noindent where MSE is defined by Equation~\ref{eqn:mse} and max(|$\delta T_b(\nu)$|) is the signal amplitude.

The distribution of mean relative rms error for all 20 trials is shown in Figure~\ref{fig:err_hist}. Across the 20 trials, {\sc 21cmLSTM} has an average relative mean error of 0.223\% $\pm$ 0.031\% (corresponding to average absolute error of 0.389 mK $\pm$ 0.047 mK), average median error of 0.197\% $\pm$ 0.025\%, and average maximum error of 0.824\% $\pm$ 0.183\%. The best trial (top right panel of Figure~\ref{fig:big}) has mean relative error of 0.18\% (corresponding to absolute error of 0.30 mK), median error of 0.16\% (corresponding to 0.26 mK), and maximum error of 0.75\% (corresponding to 1.34 mK). Therefore, when trained and tested on the same data for the same number of trials, {\sc 21cmLSTM} has a $1.6\times$ lower average error and $\approx2\times$ lower maximum error than those reported for {\sc 21cmVAE} (see Table~\ref{tab:results}).

When trained and tested on the equivalent {\tt ARES} sets, the best trial (bottom right panel of Figure~\ref{fig:big}) has mean relative error of 0.29\% (corresponding to 0.42 mK), median error of 0.21\%, and max error of 1.77\% (see Section~\ref{subsec:model_comparison}).

\begin{table}
    \caption{Accuracy and Speed Metrics of Global 21 cm Signal Emulators}
    \begin{tabular}{cccc}
    \toprule
    \addlinespace
    Emulator& Mean Error& Maximum Error & Speed\\
    &(\%)& (\%)& (ms)\\
    \addlinespace
    \hline
    {\sc 21cmLSTM} & 0.22 & 0.82 & 46\\
    {\sc 21cmVAE} & 0.35 & 1.84 & 74\\
    {\tt globalemu} & 1.12 & 6.32 & 3\\
    {\sc 21cmGEM} & 1.59 & 10.55 & 160\\
    \bottomrule
    \end{tabular}
    \begin{tablenotes}
    \small
    \item The information provided for each emulator are the average mean and maximum rms errors across the full frequency range of $\approx$1700 test signals (see Section~\ref{subsec:accuracy}), and the average evaluation speed when predicting one signal at a time (see Section~\ref{subsec:speed}). The errors quoted for other emulators are from their respective original paper \citep{Cohen20,globalemu,21cmVAE}. For direct comparison purposes, the speeds quoted for the first three emulators were measured using the same computational resources stated in Section~\ref{subsec:training}, while we note that the speeds measured in the original papers for {\sc 21cmVAE} \citep{21cmVAE} and {\tt globalemu} \citep{globalemu} are 41.4 ms and 1.3 ms, respectively. The speed quoted for {\sc 21cmGEM} is from its original paper \citep{Cohen20}.
    \label{tab:results}
    \end{tablenotes}
\end{table}

\subsection{Speed} \label{subsec:speed}
We report the emulation speed of {\sc 21cmLSTM} as the average time to predict a single global 21 cm signal in the {\sc 21cmGEM} test set (i.e., predict $\delta T_b$ for all $n=451$ frequencies) from the seven input physical parameters, or the emulator evaluation time including steps for data preprocessing (see Section~\ref{subsec:data}) and signal denormalization (see Equation~\ref{eqn:minmax}).~We employed the same computational resources stated in Section~\ref{subsec:training} and used the {\tt time} module to measure the total processing time, which we note naturally depends on the computing power (e.g., number and type of CPU cores and GPUs). We report the speed for a single evaluation\footnote[10]{This uses the {\tt eval\_21cmGEM.py} GitHub script (see first footnote link).} for proper benchmarking with other emulators, as some architectures are inherently more conducive to parallel processing than RNNs, which are serial in nature.

Across 20 trials, the average emulation speed of {\sc 21cmLSTM} is 46 ms, which is similar to those reported for other emulators of the global 21 cm signal, except for {\tt globalemu} \citep{globalemu} which was designed to be faster. We performed the same timing test using the latest versions of {\tt 21cmVAE} \citep{21cmVAE} and {\tt globalemu} \citep{globalemu} and measured their average speeds to be 74 ms and 3 ms, respectively (see Table~\ref{tab:results}). The speed and unprecedented accuracy of {\sc 21cmLSTM} make it capable of efficient Bayesian multi-parameter estimation, as we carry out in Section~\ref{sec:posterioremulation}, which reflects the success of a two-layer LSTM RNN in emulating 21 cm models.

\subsection{Model Comparison} \label{subsec:model_comparison}
In Section~\ref{subsec:accuracy}, we found that {\sc 21cmLSTM} performs somewhat better when trained and tested on the {\sc 21cmGEM} data sets than on the {\tt ARES} sets, with a best trial mean relative error of 0.183\% compared to 0.288\% (right panel of Figure~\ref{fig:big}). We remind the reader that the {\sc 21cmGEM} data set was created by a large-volume seminumerical model \citep{Visbal12, Fialkov13, Fialkov14} similar to {\sc 21cmFAST} (\citealt{Mesinger11}; see \citealt{Cohen20}), while {\tt ARES} is a semianalytical model that does not calculate 3D volumes. The difference in performance is likely caused by differences in the parameterizations and parameter ranges between the models, which can be qualitatively compared in Table~\ref{tab:params}. In particular, a single parameter for the SFE ($f_{\star}$) is varied in the {\sc 21cmGEM} sets, while four SFE parameters ($f_{\rm \star,0}$, $M_{\rm p}$, $\gamma_{\rm lo}$, $\gamma_{\rm hi}$, which describe a double power-law) are varied in the {\tt ARES} sets, which may result in a smaller range of cosmic star formation histories and thus less variation among the signals in the {\sc 21cmGEM} sets compared to {\tt ARES} (see left panel of Figure~\ref{fig:big}).

\begin{figure}
    \includegraphics[scale=0.52]{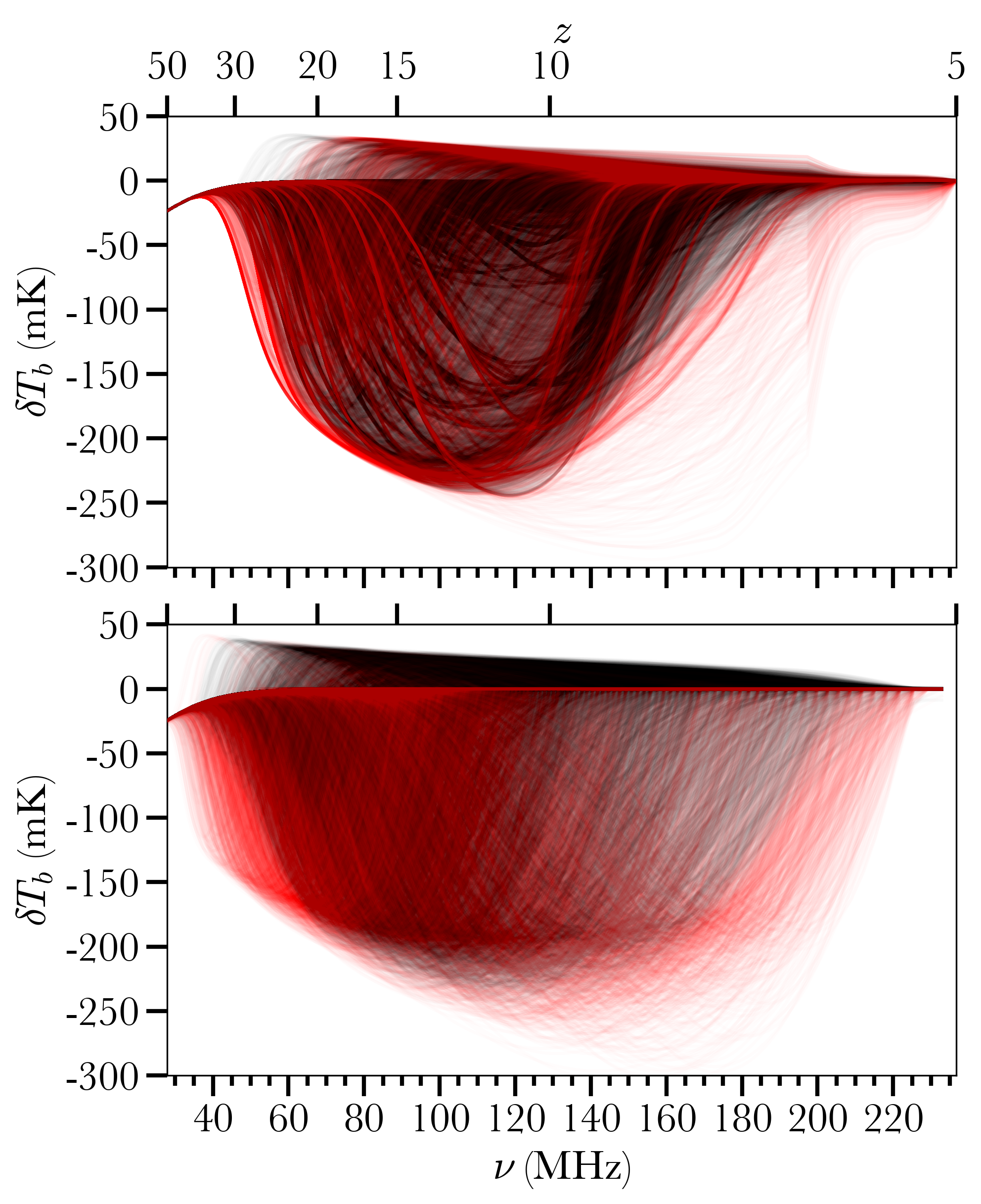} \caption{{\it Top:} 15,000 signals randomly selected from {\sc 21cmGEM} combined training+validation set with PCA outliers shown in red (see Section~\ref{subsec:model_comparison}). {\it Bottom:} Same as the top panel but for {\tt ARES}.} \label{fig:PCA}
\end{figure}

We briefly investigated the differences between the two combined training+validation sets by performing a Principal Component Analysis (PCA) decomposition of each set. By default, we set the number of components extracted equal to the number of features, or the number of physical parameters varied in each data set (i.e., seven for {\sc 21cmGEM} and eight for {\tt ARES}). We calculate the Mahalanobis distance \citep{mahalanobis18} for each signal, which is a common metric used for multidimensional outlier detection and defined between two points $u$ and $v$ as $d=\sqrt{(u-v)(1/V)(u-v)^T}$, where $(1/V)$ is the inverse covariance. We define outliers as those signals with $d>3$, meaning they are $>3\sigma$ from the sample mean vector. We find that 32\% (8499) of the {\tt ARES} set are outliers, while 25\% (6800) of the {\sc 21cmGEM} set are outliers (shown in red in Figure~\ref{fig:PCA}). This statistical analysis is consistent with larger variation in the {\tt ARES} set causing {\sc 21cmLSTM} to have a higher emulation error when trained and tested on {\tt ARES} compared to the {\sc 21cmGEM} set. Beyond the visual comparison of model features offered by Figure~\ref{fig:PCA}, we leave for future work a detailed study of the similarities and differences between these two popular models of the global 21 cm signal.

\section{Posterior Emulation} \label{sec:posterioremulation}
In this section, we use {\sc 21cmLSTM} as the model in the likelihood of Bayesian nested sampling analyses to fit mock global 21 cm signals with added statistical noise and numerically estimate seven astrophysical parameters. We describe the steps to obtain converged posterior distributions, and we present the signal posterior constraints obtained by the emulator compared to the fiducial signal for three 21 cm noise levels. Note that this analysis ignores systematic uncertainties from the beam-weighted foreground (see, e.g., \citealt{Bernardi16, Hibbard20, Anstey23, Sims23, Hibbard23, Pagano24, Saxena24}), RFI \citep{Shi22, Leeney23}, and environmental effects (see, e.g., \citealt{SARAS2, Kern20, Bassett21, Shen22, Murray22, Pattison24}).

\subsection{Bayesian Inference Analysis} \label{subsec:inference}
\begin{figure*}
    \includegraphics[scale=0.355]{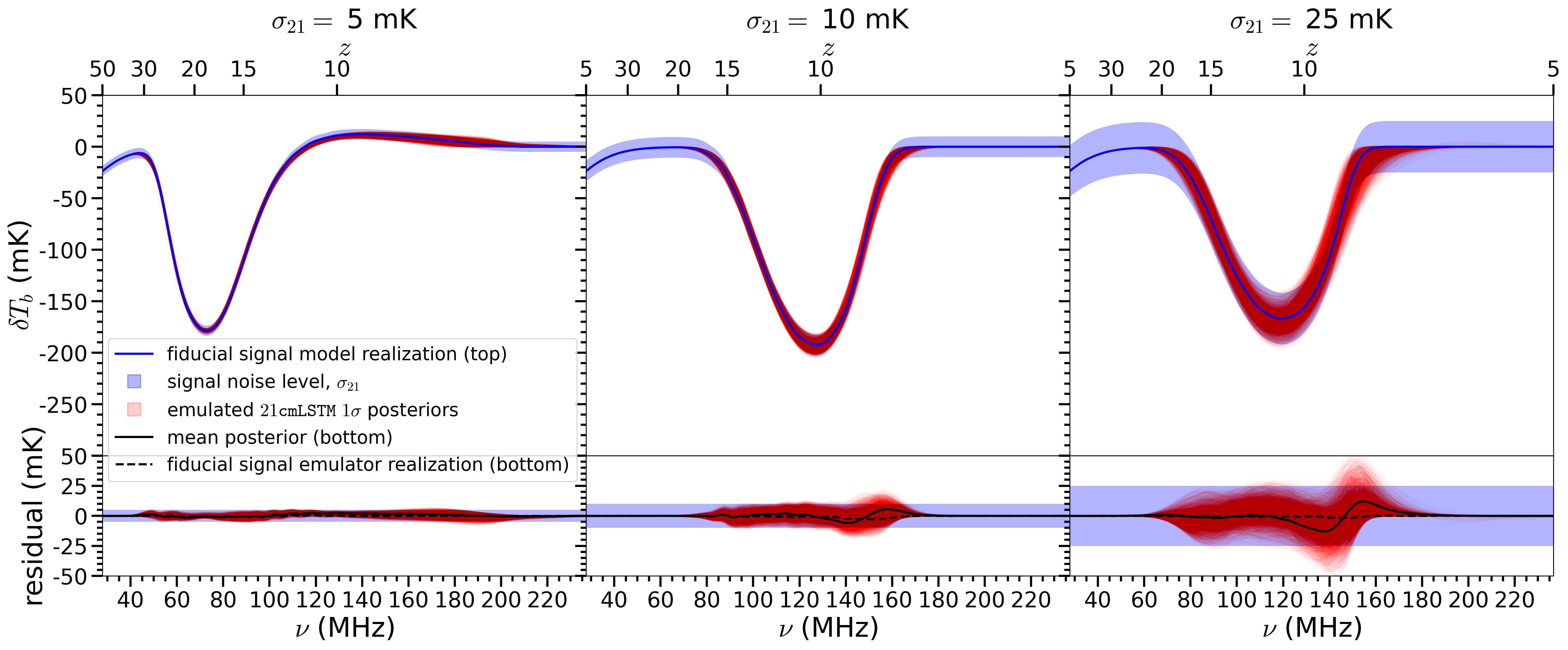} \caption{{\it Top:} Signal realizations of the $1\sigma$ posterior samples (red, see Section~\ref{subsec:posterior}) obtained from Bayesian nested sampling analyses using {\sc 21cmLSTM} to fit three fiducial global 21 cm signals (dark blue) randomly selected from the {\sc 21cmGEM} test set with added Gaussian-distributed noise (light blue bands) of 5 mK (left), 10 mK (middle), and 25 mK (right). {\it Bottom:} Residuals between the corresponding true fiducial signal and each {\sc 21cmLSTM} $1\sigma$ posterior (red, see Table~\ref{tab:multinest}), the posterior mean (solid black), and the emulation of the fiducial signal (dashed black).}
    \label{fig:posterior}
\end{figure*}

\begin{table}
    \centering
    \caption{Summary of Nested Sampling Analyses\label{tab:multinest}}
    \begin{tabular}{cccccc}
    \toprule
    \addlinespace
        $\sigma_{21}$&$n_{\rm live}$&$n_{\rm evaluations}$&$f_{\rm accept}$&$\log {\it Z}$&$\sigma_{\rm posterior}$\\
        (mK)& & & & &(mK)\\
    \addlinespace
    \hline
    5&1200&125,232&0.202&-255.9$\pm$0.1&1.6\\
    10&1200&64,106&0.340&-254.6$\pm$0.1&2.9\\
    25&1200&46,784&0.369&-251.1$\pm$0.1&7.9\\
    \bottomrule
    \end{tabular}
    \begin{tablenotes}
    \small
    \item The information provided for each fit are the noise level of the mock 21 cm signal ($\sigma_{21}$), the number of initial live points ($n_{\rm live}$), the total number of likelihood evaluations ($n_{\rm evaluations}$), the final acceptance rate ($f_{\rm accept}$), the final evidence ($\log {\it Z}$), and the posterior $1\sigma$ rms error ($\sigma_{\rm posterior}$). The analysis methods and results are described in Section~\ref{subsec:inference} and Section~\ref{subsec:posterior}, respectively. The $1\sigma$ posterior signal realizations and residuals with respect to the true fiducial signal are shown in Figure~\ref{fig:posterior}, and the full posterior distributions for $\sigma_{21}=$ 25 mK and $\sigma_{21}=$ 5 mK are shown in Figure~\ref{fig:corner_25mK} and Figure~\ref{fig:corner_5mK}, respectively. 
    \end{tablenotes}
\end{table}

We perform Bayesian parameter inference analyses to numerically estimate the posterior distribution $P(\theta | \boldsymbol{D}, m)$ of a set of parameters $\theta$ in a physical model $m$, given observed (mock) data $\boldsymbol{D}$ with priors $\pi$ on the parameters. Bayes' theorem states this as:
\begin{equation} \label{eqn:Bayes}
P(\theta | \boldsymbol{D}, m) = \frac{\mathcal{L} (\theta) \pi (\theta)}{{\it Z}},
\end{equation}

\noindent where $\mathcal{L}$ is the likelihood function and {\it Z} is the Bayesian evidence, which can be used for model comparison. We sample from a multivariate log-likelihood function assuming Gaussian distributed noise:
\begin{equation} \label{eqn:likelihood}
\log \mathcal{L}(\theta) \propto [\boldsymbol{D} - m(\theta)]^T \boldsymbol{C}^{-1} [\boldsymbol{D} - m(\theta)],
\end{equation}

\noindent where the noise covariance, $\boldsymbol{C}$, is a diagonal array of constant values corresponding to the square of the estimated noise $\sigma_{21}$.

We employ the Bayesian inference method of nested sampling (\citealt{Skilling04}; for reviews see \citealt{Ashton22}; \citealt{Buchner23}), which converges on the best parameter estimates by iteratively removing regions of the prior volume with lower likelihood and computes the evidence and posterior samples simultaneously. As mentioned in the introduction, Monte Carlo methods like nested sampling and MCMC are computationally expensive because they require many likelihood evaluations to sample the multi-dimensional posterior, and so model emulators are desired to speed up or make feasible such analyses. We choose nested sampling rather than MCMC because the former is designed to constrain parameter spaces with complex degeneracies or multimodal distributions \citep{Buchner23}, which are expected when fitting 21 cm mock or real data (e.g., \citealt{Bevins22b, DorigoJones23, Breitman24}; also see \citealt{Saxena24}). For all analyses, we employ {\tt MultiNest} (\citealt{Feroz08, Feroz09, Feroz19}) with default evidence tolerance and sampling efficiency and $3\times$ the default initial ``live'' point number (1200), which we find results in consistent, converged posteriors. For an in-depth description of {\tt MultiNest} and other algorithms, see, e.g., \citet{Lemos23}.

We use {\sc 21cmLSTM} trained on the {\sc 21cmGEM} set as the model for the global 21 cm signal in the likelihood. The trained instance of {\sc 21cmLSTM} used for all analyses has mean rms error of 0.20\% (corresponding to 0.33 mK) and maximum error of 0.63\% when evaluated on the {\sc 21cmGEM} test set, which is consistent with the average accuracy found in Section~\ref{subsec:accuracy}. We evaluate this emulator's ability to constrain a synthetic 21 cm signal with added statistical noise.

We fit three different fiducial mock global 21 cm signals that are randomly selected from the {\sc 21cmGEM} test set and have different levels of added statistical white noise that is Gaussian-distributed. The three different 21 cm noise levels tested include the optimistic and fiducial scenarios for the REACH radiometer \citep{REACH}: $\sigma_{21}=5$ mK or 10 mK (referred to as ``optimistic''), and $\sigma_{21}=25$ mK (referred to as ``standard''), where $\sigma_{21}$ is the standard deviation noise estimate. We note that in \citet{DorigoJones23}, we tested these same noise levels, as well as 50 mK and 250 mK, and compared emulated posteriors obtained using {\tt globalemu} \citep{globalemu} to the corresponding ``true'' posteriors obtained using {\tt ARES}. From the ideal radiometer sensitivity equation (e.g., \citealt{Kraus66}) for a non-systematics-limited 21 cm experiment, assuming $\nu=$30 MHz (i.e., Dark Ages) and $\Delta\nu=$0.5 MHz, the noise levels 5 mK, 10 mK, and 25 mK correspond to integration times of $\approx7100$ hours, $\approx1800$ hours, and $\approx300$ hours, respectively.

\subsection{Posterior Results} \label{subsec:posterior}
In the top panels of Figure~\ref{fig:posterior}, we present sets of posterior signal realizations (shown in red), when using {\sc 21cmLSTM} to fit three different mock 21 cm signals (shown in dark blue) with added noise levels (shown in light blue) of 5 mK, 10 mK, and 25 mK, from left to right. We show the $1\sigma$ posteriors in red, defined as the 68\% of samples with the lowest relative rms error with respect to the fiducial signal (Equation~\ref{eqn:rel_error}). In the bottom panels of Figure~\ref{fig:posterior}, we present the residuals between the fiducial signal and each $1\sigma$ posterior sample (i.e., $\delta T_b(\nu)-\hat{\delta T_b}(\nu)$; red), the mean of all the posterior samples (solid black), and the emulator realization of the fiducial signal (i.e., $m_{{\tt 21cmLSTM}}(\theta_0)$; dashed black). Table~\ref{tab:multinest} summarizes each fit.

We find that, for each 21 cm noise level tested, the posterior mean residual is significantly less than the signal noise estimate, $\sigma_{21}$, across the full redshift range and approaches the emulator error (0.33 mK, see Section~\ref{subsec:inference}) as the noise level decreases. This is seen visually in the bottom panels of Figure~\ref{fig:posterior}, as well as quantitatively in the mean and $1\sigma$ (i.e., 68th percentile) rms errors of the posteriors for each fit (see $\sigma_{\rm posterior}$ in Table~\ref{tab:multinest}). The mean relative rms error (Equation~\ref{eqn:rel_error}) between all the emulated posteriors and the true fiducial signal is 0.87\% (corresponding to 1.56 mK absolute error) for $\sigma_{21}=5$ mK, 1.42\% (corresponding to 2.73 mK) for $\sigma_{21}=10$ mK, and 4.11\% (corresponding to 6.87 mK) for $\sigma_{21}=25$ mK. The posterior mean and $1\sigma$ errors are thus each $\approx3\times$ less than $\sigma_{21}$ for each fit. The fit obtained using {\sc 21cmLSTM} consistently improves for decreasing 21 cm noise levels, corresponding to longer integration times, as expected due to the increase in constraining power. We note that this general trend of a more accurate fit to the mock signal for decreasing noise levels is robust as it does not depend on the random signals being fit.

In addition to the posterior signal realizations discussed, we can examine the marginalized 1D and 2D posterior distributions. We present the full posterior parameter distribution for the $\sigma_{21}=25$ mK fit in Figure~\ref{fig:corner_25mK} and for the $\sigma_{21}=5$ mK fit in Figure~\ref{fig:corner_5mK}. For the standard noise level tested (i.e., $\sigma_{21}=25$ mK), we find that the 1D posteriors for three astrophysical parameters ($f_*$, $V_c$, and $\tau$) are well-constrained and unbiased with respect to (i.e., within $2\sigma$ of) their fiducial values, while the other four parameters ($f_X$, $\alpha$, $\nu_{\rm min}$, and $R_{\rm mfp}$) are relatively unconstrained. For the optimistic noise levels tested (i.e., $\sigma_{21}=5$ mK and 10 mK), our findings are similar, although the constraints improve somewhat, in particular for $f_X$ and $\nu_{\rm min}$, reflecting the improved signal posterior realizations seen in Figure~\ref{fig:posterior}.

These results demonstrate that, as a result of its low emulation error, {\sc 21cmLSTM} can sufficiently exploit even outstandingly optimistic measurements of the global 21 cm signal and obtain unbiased posterior constraints. We find that, from our non-systematics-limited global 21 cm mock data analysis, we obtain unbiased posterior constraints when the emulator error is $\approx1\%-5\%$ of the signal observational noise level, $\sigma_{21}$. These ratio values are based on the emulation rms errors for the fiducial mock signals (black dashed lines in bottom panels of Figure~\ref{fig:posterior}) fit with $\sigma_{21}=25$ mK and $\sigma_{21}=5$ mK, which are 0.31 mK and 0.24 mK, respectively. Furthermore, these results are consistent with other comparable Bayesian analyses of mock 21 cm data that have found jointly fitting complementary summary statistics or data sets is needed to break the degeneracies between certain astrophysical parameters (e.g., \citealt{Qin20, Chatterjee21, Bevins23, DorigoJones23, Breitman24}).

\section{Conclusions} \label{sec:conclusions}
Achieving unbiased Bayesian parameter inference of the global 21 cm signal using a neural network emulator requires the emulation error to be much lower than the observational noise on the signal (e.g., \citealt{DorigoJones23}). Highly accurate and fast emulation is therefore needed to sufficiently exploit optimistic or standard measurements of the 21 cm signal, especially to approach the cosmic variance limit of $\sim$0.1 mK in the future \citep{Munoz21}. To this end, in this paper, we presented a new emulator of the global 21 cm signal, called {\sc 21cmLSTM}, which is a long short-term memory recurrent neural network that has exceptionally low emulation error compared to existing emulators, which are all fully connected neural networks. {\sc 21cmLSTM} owes its unprecedented accuracy to its unique ability to leverage the intrinsic (spatiotemporal) correlation of information between neighboring frequency channels in the global 21 cm signal.

In Section~\ref{sec:methods}, we optimized {\sc 21cmLSTM} by testing different architectures (i.e., number of LSTM layers, activation functions, loss function, Bi-LSTM models), data preprocessing steps (i.e., normalizations), and training configurations (i.e., number of epochs and batch sizes). A schematic diagram of the network architecture of {\sc 21cmLSTM} is shown in Figure~\ref{fig:emulator}. In Section~\ref{sec:results}, we presented the emulation accuracy of {\sc 21cmLSTM} when trained and tested on large data sets created by two different, popular models of the global 21 cm signal (see Figure~\ref{fig:big}). Finally, in Section~\ref{sec:posterioremulation}, we employed a representative instance of {\sc 21cmLSTM}, trained on a {\sc 21cmGEM} set, as the model in the likelihood of a Bayesian nested sampling analysis to fit mock signals and showed that it can be used to obtain unbiased posterior constraints.

When trained and tested on the same data as existing emulators, {\sc 21cmLSTM} has an average relative rms error of (0.22$\pm$0.03) \% (Figure~\ref{fig:err_hist}), corresponding to (0.39$\pm$0.05) mK, and best trial mean error of 0.18 \% (top right panel of Figure~\ref{fig:big}), corresponding to 0.30 mK. {\sc 21cmLSTM} therefore has a $\approx1.6\times$ lower average error than the previously most accurate emulator of the global 21 cm signal, {\tt 21cmVAE} (Table~\ref{tab:results}). The maximum emulation error of {\sc 21cmLSTM} is 0.82\% on average, which is $\approx2\times$ lower than that reported for {\tt 21cmVAE}. Furthermore, {\sc 21cmLSTM} has similar emulation speed as other existing emulators when predicting one signal at a time (Table~\ref{tab:results}, Section~\ref{subsec:speed}), making it both sufficiently fast and accurate for complex, high-dimensional Bayesian parameter estimation analyses. We also examined a set of 21 cm signals created by the {\tt ARES} model with a greater parameter variation than the {\sc 21cmGEM} set (Figure~\ref{fig:PCA}, Section~\ref{subsec:model_comparison}) and found, as it might be expected, that {\sc 21cmLSTM} produces a somewhat higher emulation error when trained on this {\tt ARES} set (bottom right panel of Figure~\ref{fig:big}).

We obtained accurate posterior distributions when using {\sc 21cmLSTM} in {\tt MultiNest} analyses to fit mock global 21 cm signals with added observational noise levels of $\sigma_{21}=5$ mK, $\sigma_{21}=10$ mK, and $\sigma_{21}=25$ mK. The full parameter posterior distributions for the $\sigma_{21}=25$ mK and $\sigma_{21}=5$ mK fits are presented in Figure~\ref{fig:corner_25mK} and Figure~\ref{fig:corner_5mK}, respectively, and the posterior signal realizations and residuals for all fits are shown in Figure~\ref{fig:posterior}. The posteriors provide a good fit to each fiducial mock signal, with the posterior mean and $1\sigma$ errors being $\approx3\times$ less than the respective, adopted signal noise level, $\sigma_{21}$ (see bottom panel of Figure~\ref{fig:posterior}, Table~\ref{tab:multinest}, Section~\ref{subsec:posterior}). The posterior mean residual consistently decreases as the signal noise level decreases, with the $\sigma_{21}=5$ mK fit having a posterior mean relative rms error of only 0.87\% (corresponding to 1.56 mk), compared to the pure emulation average error of 0.20\% (corresponding to 0.33 mK) for the instance of {\sc 21cmLSTM} employed. For all three noise levels tested, the posterior distributions are well-converged and unbiased for three of seven parameters ($f_*$, $V_c$, and $\tau$), and for the lowest noise level (i.e., $\sigma_{21}=5$ mK), the posteriors become unbiased for two more parameters ($f_X$ and $\nu_{\rm min}$). These results are consistent with recent findings that jointly fitting complementary summary statistics or data sets is needed to constrain certain astrophysical parameters (e.g., \citealt{Qin20, Chatterjee21, Bevins23, DorigoJones23, Breitman24}).

This work demonstrates that LSTM recurrent neural networks exploit the intrinsic correlation of adjacent frequency channels (i.e., autocorrelation) in the global 21 cm signal to perform very accurate and fast emulation of physically-motivated seminumerical or semianalytical models of the signal. We have made the data sets and code publicly available on Zenodo\footnoteref{data} and GitHub\footnoteref{code}, respectively, so that the {\sc 21cmLSTM} emulator can be used and modified by the community. In principle, {\sc 21cmLSTM} could also be adapted to learn the pattern of and predict any sequential or time series measurement, assuming sufficient data size and resolution, and subsequently be employed in Bayesian analyses. The publicly available emulator {\sc 21cmLSTM} contributes to the growing body of astrophysics and cosmology research finding that, for data or measurements with intrinsic correlation over time, LSTM recurrent neural networks can perform as well as or better than fully connected neural networks.

\begin{acknowledgments}
We thank the anonymous reviewer for their thorough feedback that improved the manuscript. We thank Harry T. J. Bevins, Christian H. Bye, and Joshua J. Hibbard for useful discussions. This work utilized the Blanca condo computing resource at the University of Colorado Boulder. Blanca is jointly funded by computing users and the University of Colorado Boulder. This work was directly supported by the NASA Solar System Exploration Research Virtual Institute cooperative agreement 80ARC017M0006. We acknowledge support by NASA APRA grant award 80NSSC23K0013 and a subcontract from UC Berkeley (NASA award 80MSFC23CA015) to the University of Colorado (subcontract \#00011385) for science investigations involving the LuSEE-Night lunar far side mission.~J.M. was supported by an appointment to the NASA Postdoctoral Program at the Jet Propulsion Laboratory/California Institute of Technology, administered by Oak Ridge Associated Universities under contract with NASA. Part of this work was done at Jet Propulsion Laboratory, California Institute of Technology, under a contract with the National Aeronautics and Space Administration (80NM0018D0004).
\end{acknowledgments}

\software{This research relies heavily on the {\sc Python} \citep{python} open source community libraries {\sc numpy} \citep{numpy}, {\sc matplotlib} \citep{matplotlib}, {\sc scipy} \citep{scipy}, {\sc tensorflow} \citep{tensorflow}, and {\sc Keras} \citep{keras}. This research also utilized {\sc jupyter} \citep{jupyter}, {\tt MultiNest} \citep{Feroz09, Feroz19}, {\tt 21cmVAE} \citep{21cmVAE}, and {\tt globalemu} \citep{globalemu}.}

\appendix \renewcommand\thefigure{\thesection.\arabic{figure}} \setcounter{figure}{0}
\section{Loss Curves for Validation and Training Sets} \label{sec:loss}
\begin{figure*}[t]
    \includegraphics[width=\textwidth]{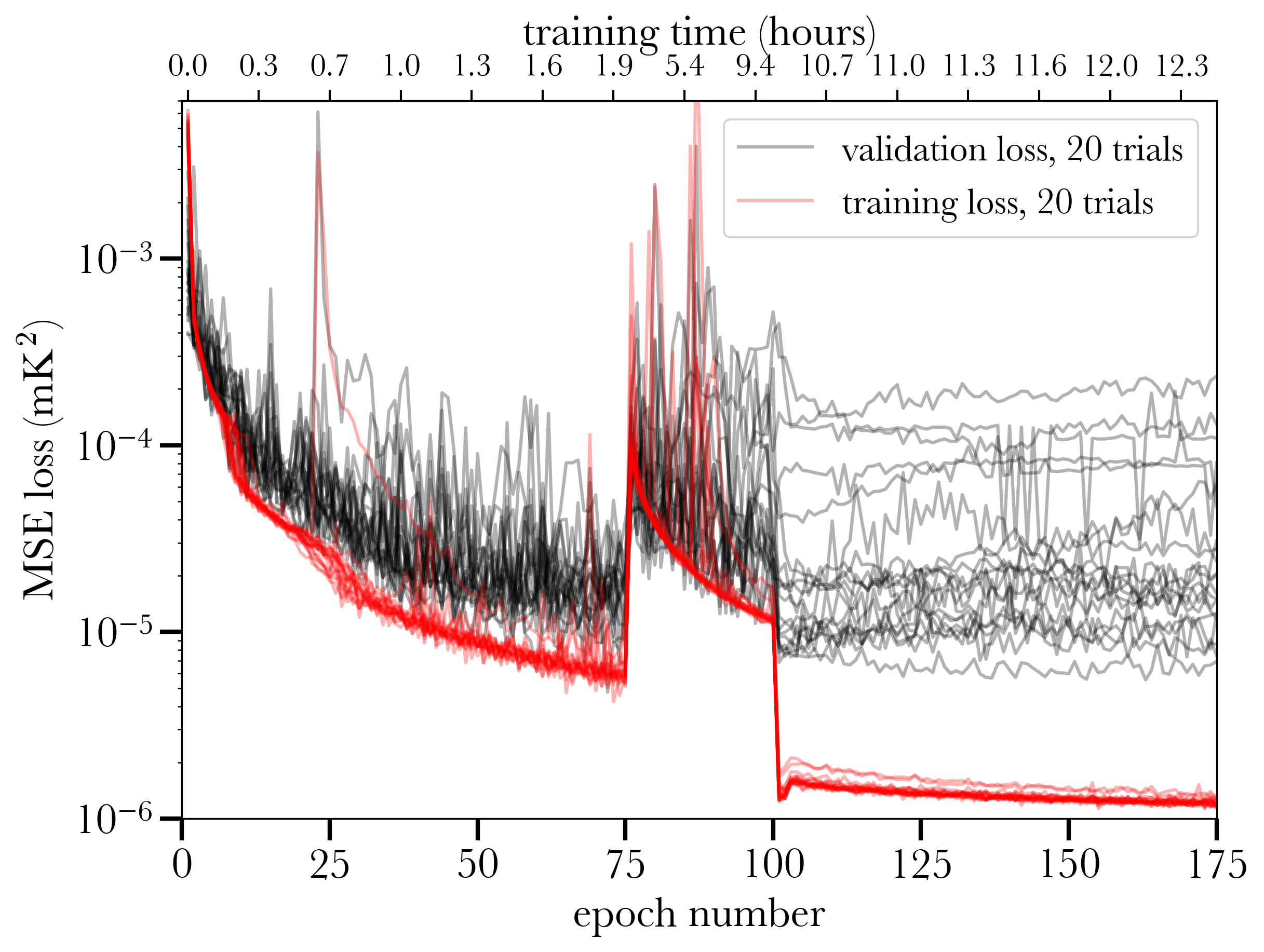} \caption{Loss versus training epoch number for validation (black) and training (red) sets for 20 trials of {\sc 21cmLSTM} trained on the {\sc 21cmGEM} data. The top axis shows the approximate training time at each epoch. Note that the emulator is trained for 75 epochs with batch size of 10 before and after training for 25 epochs with batch size of one (see batch scheduling description in Section~\ref{subsec:training}).}
    \label{fig:loss}
\end{figure*}

Figure~\ref{fig:loss} shows the distribution of MSE loss (Equation~\ref{eqn:mse}) for the training and validation sets for 20 identical trials of {\sc 21cmLSTM} trained on the {\sc 21cmGEM} data (see Section~\ref{subsec:training} and Section~\ref{subsec:accuracy}). The validation loss for each trial reaches a stable value, which indicates that the network is able to generalize to unseen signals and is not overfitting the training set, whereas increasing validation loss would indicate overfitting.

\setcounter{figure}{0}
\section{Posterior distributions from fitting mock global 21 cm signals with 25 mK and 5 mK noise} \label{sec:app}
\begin{figure*}[t]
    \includegraphics[width=\textwidth]{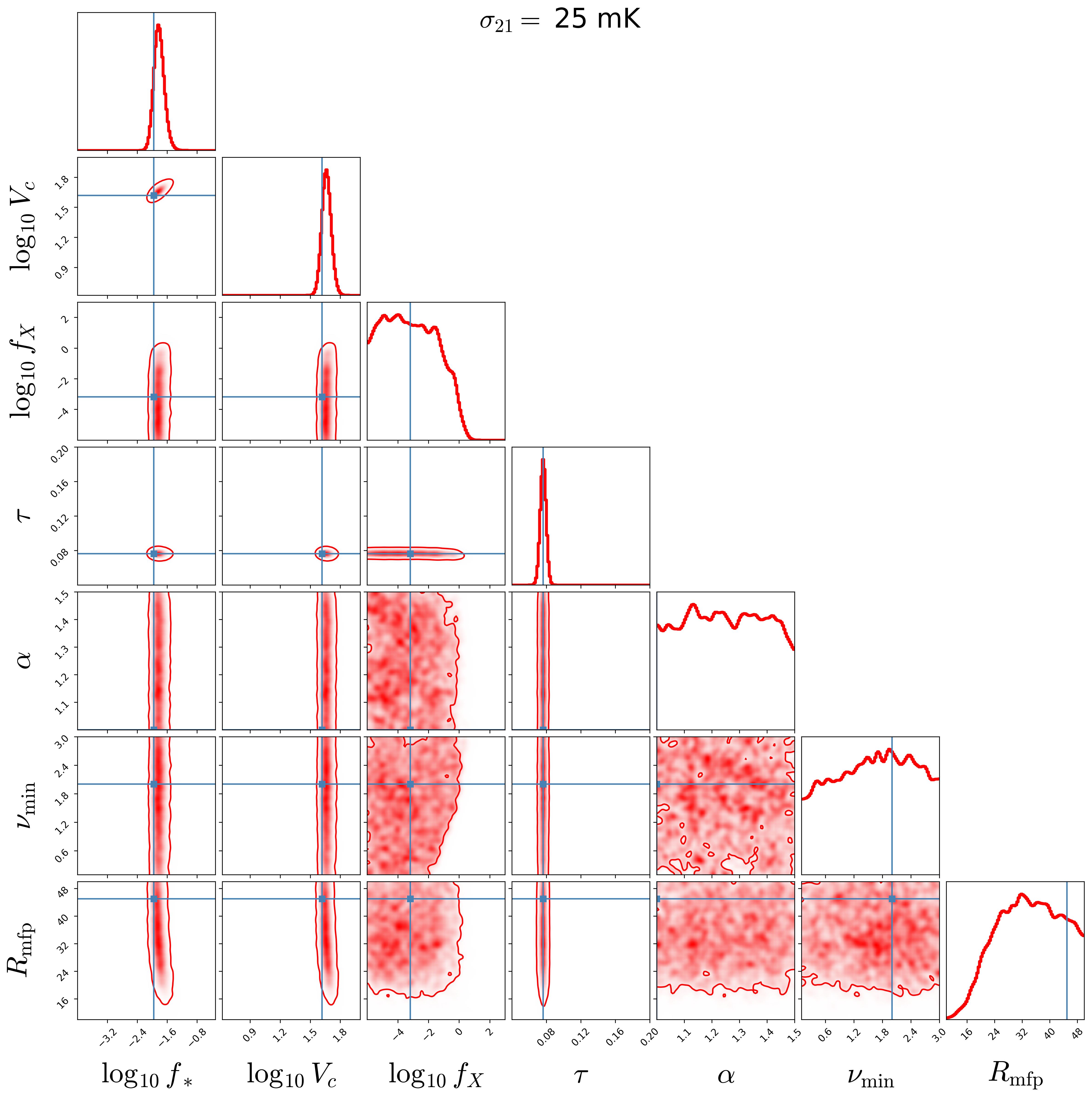} \caption{Marginalized 1D and 2D posterior distributions for the seven astrophysical parameters of the {\sc 21cmGEM} set obtained when using {\sc 21cmLSTM} to fit global 21 cm signal mock data with observational noise of $\sigma_{21}=$ 25 mK (see Section~\ref{sec:posterioremulation}). These parameters control the CMB optical depth, SFE, and UV and X-ray photon production in galaxies (Table~\ref{tab:params}). Blue vertical and horizontal lines indicate the fiducial parameter values of the mock signal being fit, $\theta_0=$($f_*$, $V_c$, $f_X$, $\tau$, $\alpha$, $\nu_{\rm min}$, $R_{\rm mfp}$)=($1.102\times10^{-2}$, 41.534, $6.470\times10^{-4}$, 0.076, 1, 2, 45), which was randomly chosen from the test set (see right panel of Figure~\ref{fig:posterior}). Contour lines in the 2D histograms represent the 95\% confidence levels, and density colormaps are shown. Axis ranges are the full prior ranges given in Table~\ref{tab:params}.}
    \label{fig:corner_25mK}
\end{figure*}

\begin{figure*}[t]
    \includegraphics[width=\textwidth]{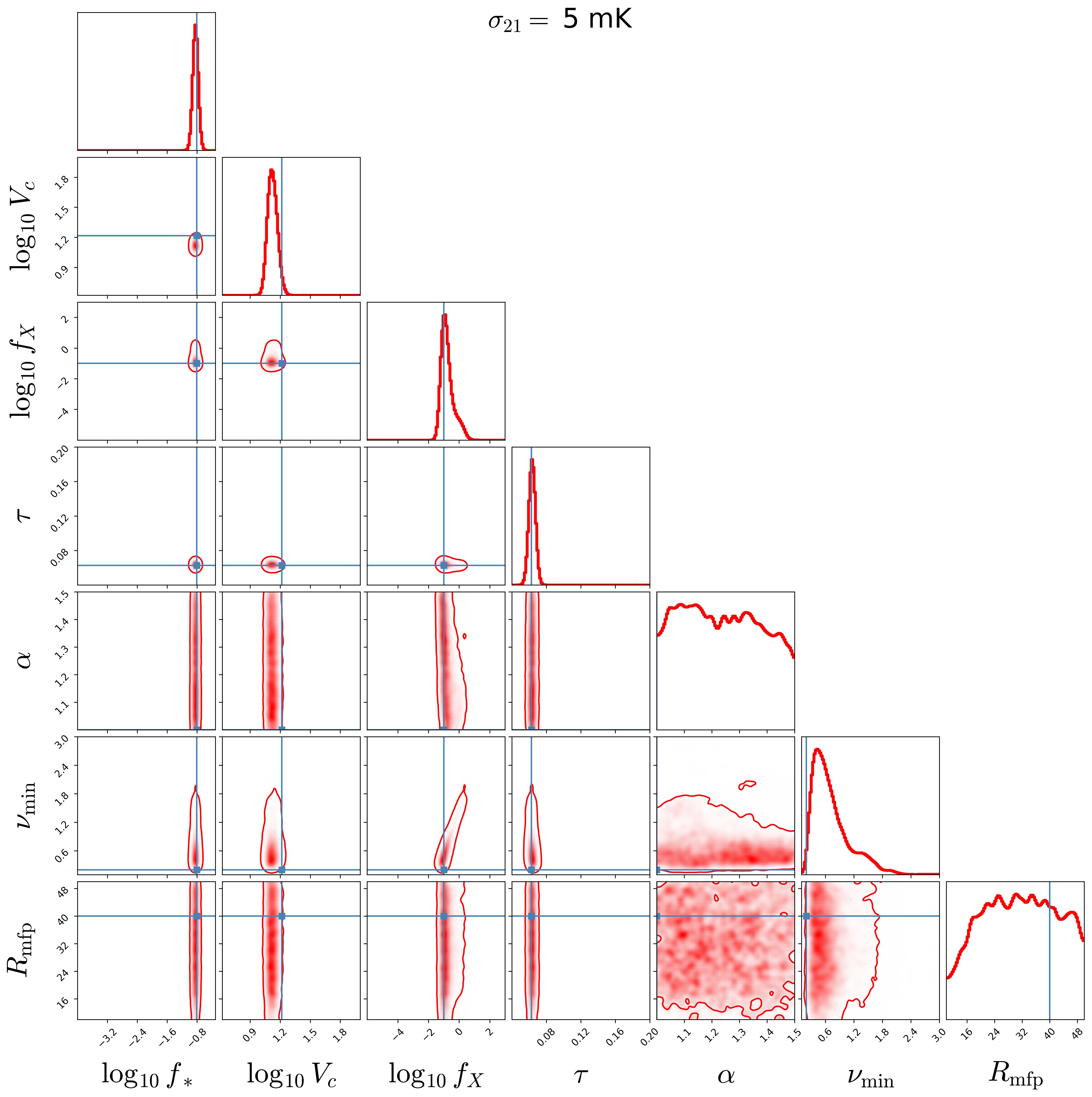} \caption{Same as Figure~\ref{fig:corner_25mK} but when using {\sc 21cmLSTM} to fit a different randomly selected signal from the {\sc 21cmGEM} test set with $\theta_0=$($f_*$, $V_c$, $f_X$, $\tau$, $\alpha$, $\nu_{\rm min}$, $R_{\rm mfp}$)=($1.581\times10^{-1}$, 16.5, 0.1, 0.0626, 1, 0.2, 40) and observational noise of $\sigma_{21}=$ 5 mK (see left panel of Figure~\ref{fig:posterior}).}
    \label{fig:corner_5mK}
\end{figure*}

In Figure~\ref{fig:corner_25mK}, we present the full 1D and 2D marginalized posterior distribution for seven astrophysical parameters obtained when using {\sc 21cmLSTM} in a nested sampling analysis to fit a mock global 21 cm signal randomly selected from the {\sc 21cmGEM} test set with standard observational noise level of $\sigma_{21}=$ 25 mK. In Figure~\ref{fig:corner_5mK}, we present the posterior distribution when fitting a different randomly selected signal fit with $\sigma_{21}=$ 5 mK. The $1\sigma$ posterior signal realizations for each fit are shown in the respective panel of Figure~\ref{fig:posterior}. For each noise level tested, {\sc 21cmLSTM} obtains unbiased posteriors for $f_*$, $V_c$, and $\tau$, and for the optimistic noise levels tested (i.e., $\sigma_{21}=5$ mK and 10 mK) the constraints improve for $f_X$ and $\nu_{\rm min}$. See Section~\ref{sec:posterioremulation} for further details.

\bibliography{dj24}{} \bibliographystyle{aasjournal}
\end{document}